# An ensemble approach to improved prediction from multitype data[*]

### Jennifer Clarke[1] and David Seo[2]

*University of Miami School of Medicine*

**Abstract:** We have developed a strategy for the analysis of newly available binary data to improve outcome predictions based on existing data (binary or non-binary). Our strategy involves two modeling approaches for the newly available data, one combining binary covariate selection via LASSO with logistic regression and one based on logic trees. The results of these models are then compared to the results of a model based on existing data with the objective of combining model results to achieve the most accurate predictions. The combination of model predictions is aided by the use of support vector machines to identify subspaces of the covariate space in which specific models lead to successful predictions. We demonstrate our approach in the analysis of single nucleotide polymorphism (SNP) data and traditional clinical risk factors for the prediction of coronary heart disease.

## Contents

## 1. Introduction

In applied research contexts the statistician is often faced with newly available data which may provide information relevant to a recently completed analysis. This sce-

---

[*]Supported by NIH Grant 5K25CA111636-03.

[1]Department of Epidemiology and Public Health, University of Miami Leonard M. Miller School of Medicine, Miami, FL, USA, e-mail: JClarke@med.miami.edu

[2]Department of Medicine, Division of Cardiology, University of Miami Leonard M. Miller School of Medicine, Miami, FL, USA, e-mail: DSeo@med.miami.edu

nario is occurring more and more frequently in medical research as genomic data becomes available which may provide information relevant to the determination of disease risk, a determination that has been traditionally based on existing clinical data. There is a need for statistical approaches to variable selection and modeling which attempt to provide improved outcome predictions in such contexts by combining information from new and existing data which may be of multiple types.

We have developed one such strategy for utilizing newly available binary data to improve binary outcome predictions from an existing model based on both continuous and binary data. There are many approaches to regression and classification in the machine learning and statistical literature that would be appropriate for modeling binary data, including CART [4], MARS [15], treed models [6], and logic regression [30], to name only a few. Since our interest is specifically in single nucleotide polymorphism (SNP) data, we have chosen to model binary data with logistic regression as well as logic regression models. The logic regression models recognize the often complex interactions that exist among SNPs and attempt to model such interactions in analyzing the relationships between SNPs and outcome status. Logic regression models also perform variable selection and model construction when the number of observations, $n$, is less than the number of covariates, $p$, which is a context of particular interest to us.

Our goal is to combine all available information in generating the best outcome predictions possible. In doing so we consider several approaches which borrow ideas from the multimodel ensemble modeling literature [11]. One approach is to take a weighted average of the predictions from the existing model and the binary data model, in the spirit of Bayesian model averaging [7]. A second approach is to build a model from all available covariates and not utilize the existing model, which was built before the newer binary covariates were available. Our final approach is a two-stage approach: determine subspaces of the covariate space on which the predictions from the existing model are accurate, and utilize the predictions from a model of the newer binary covariates on the remaining subspaces. This would yield a more accurate set of predictions overall in situations where neither data type is globally informative, for example, where the data have been collected from a heterogeneous population. To avoid a subspace definition which requires knowledge of the outcome of interest for observations to be predicted, we differentiate these various subspaces via support vector machines (SVM) [3, 8, 39]. As a result our technique yields "honest" predictions for new observations.

Initially we discuss the model classes and variable selection for binary data. We then discuss how the predictions from such models can be used to improve the predictions from models based on existing data via support vector machines. Our approach is demonstrated in the context of prediction of coronary heart disease from traditional clinical risk factors and genetic (SNP) data.

## 2. Model types

We assume a continuous response variable $Y$ and a $p$-dimensional vector of binary covariates $\mathbf{X}$ (the "newly available" data). In the case of SNP data each covariate $X_j, j = 1, \ldots, p$, is binary. Since the relationship between the covariates and the response is unknown we consider two model types, logistic regression and logic regression [30]. As logistic regression is a well known modeling technique we will not discuss it in detail. However we will discuss variable selection prior to logistic regression modeling when $n < p$ in Section 2.2. Logic regression is discussed in more detail below.

## 2.1. Logic regression

Logic regression [20, 30] is an adaptive regression methodology for finding Boolean combinations of binary covariates that are associated with an outcome variable. This methodology was developed to address situations where the interaction of many predictors is responsible for differences in the response, which is often the case when all predictors are binary. As described in [30] logic regression models take the form

$$(2.1) \qquad g(E[Y]) = \beta_0 + \sum_{i=1}^{t} \beta_i L_i,$$

where $L_i$ is a Boolean expression of the covariates $X_j$. A score function relates fitted values to the response. This framework includes linear regression ($g(E[Y]) = E[Y]$ with score function $RSS$), logistic regression ($g(E[Y]) = \log(E[Y]/(1-E[Y])$ with score function binomial deviance), as well as classification ($\hat{Y} = I(L = 1)$ where $I(\cdot)$ is the indicator function and the score function is $\sum(Y \neq \hat{Y})$). Logic regression models can be conveniently represented in tree form. For example, the tree in Figure 1 represents the logic expression

$$(2.2) \qquad (((X_{79}^c) \vee ((X_{48}^c) \wedge (X_{64}^c))) \wedge (((X_{28}^c) \vee (X_9^c)) \vee ((X_{43}^c) \wedge X_{63}))).$$

where $X_j$ indicates $X_j = 1$ and $X_j^c$ indicates the conjugate ($X_j = 0$).

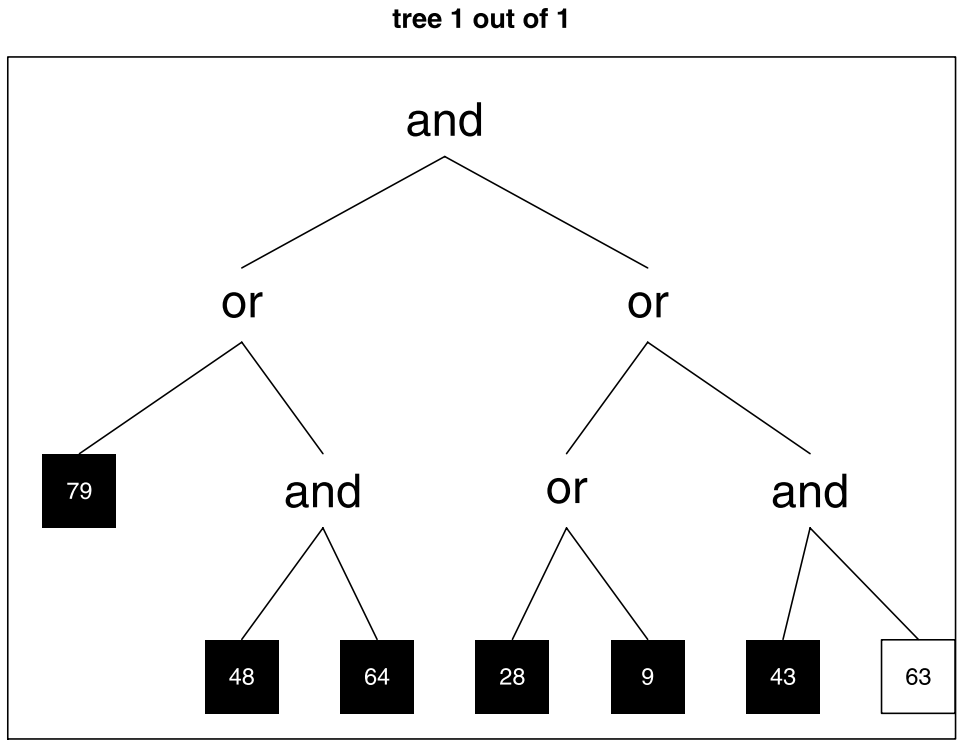

Fig 1. *A logic tree representing the Boolean expression in Equation (2.2). White text on a black background denotes the conjugate of a variable.*

Note that each $L_i$ in Equation (2.1) may be represented as a tree, and hence logic regression allows for multiple tree models.

The space of possible logic trees is enormous, especially in situations where $n < p$. To search this space efficiently without sacrificing the desire for optimality, either a greedy search or a search via simulated annealing can be employed. These search techniques estimate the $L_i$ and $\beta_i$ simultaneously (see Equation (2.1)) and use simple "moves" to search for "good" logic models (i.e., models which minimize the scoring function). Using terminology similar to that of CART [4] these "moves" include growing, pruning, splitting, and deleting. As greedy searches often lead to models which overfit the data or are suboptimal (as when the search gets "stuck" in a local minimum) [35] we prefer the use of simulated annealing to search for logic trees. Note that each "move" mentioned above has a matching "countermove" (e.g., growing as opposed to pruning) which is important in the Markov chain theory which underlies simulated annealing [38].

We use randomization to both test the null model of no signal in the data and determine the optimal model size (if the test is rejected). For testing the null model we randomly permute the response values and find the best fitting model. If there is no signal, the score of this model should be comparable to the score of the best model fit to the original data. By repeating the above procedure multiple times we can consider the number of runs with model scores better than the score of the best model fit to the original data as a p-value for our test.

The method for finding the optimal model size is based on a series of randomization tests. The null hypothesis for each test is that the optimal model size is $k$ and larger models with better scores are due to noise. Assume the null hypothesis and the best model of size $k$ has score $s_k$. The fitted values from this model fall into two classes; we now permute the response values within each class and find the best model of any size on the permuted data. If this model has score $s_k^*$ then under the null hypothesis $s_k$ comes from the same distribution as $s_k^*$. This distribution can be approximated by repeated permutations. We perform the above process for $k \in \{0, \ldots, K\}$ yielding a series of histograms of randomization scores $s_k^*$ for each value of $k$. The optimal model size is determined by comparing these histograms, for example, one may choose the model size for which only a small proportion of scores $s_k^*$ are better than $s_k$.

To further avoid overfitting the data set of $n$ observations on $p$ covariates $X_j$, $j = 1, \ldots, p$, is split into a training set of size $n_1$ and a test set of size $n_2$ ($n = n_1 + n_2$). The logic regression models are fit to the training set and the accuracy of their predictions are evaluated on the test set. The fitting and evaluation of models can be performed in the R package `LogicReg` as described in [29].

## *2.2. Variable selection*

Unlike logic trees, logistic regression models require that $n < p$. In cases where $n \geq p$ we perform a variable selection via least absolute shrinkage and selection operator (LASSO) [36] prior to regression modeling. LASSO retains the beneficial features of both subset selection and ridge regression by minimizing the residual sum of squares subject to the constraint that the sum of the absolute values of the coefficients on the covariates is less than a constant (i.e., a constraint on the $L_1$ norm of the coefficient vector). This tends to shrink some coefficients and set others to zero, leading to models with improved interpretability and stability. LASSO can be applied to generalized regression models such as logistic regression models; see [36] for details.

Osborne et al. [25] developed an efficient algorithm for computing LASSO estimates which is applicable in the $n < p$ case. We use this algorithm as implemented in the R package `lasso2` [21] in an iterative fashion to perform variable selection, removing those covariates whose coefficients has been set to zero at each iteration. If the iterative LASSO technique yields $p^* \geq n$ variables with non-zero coefficients we remove variables one at a time between LASSO iterations, starting with those variables with the smallest coefficients, until $p^* < n$. The remaining variables are used in developing a logistic regression model of our response variable via stepwise selection.

It is important to mention that variable selection techniques exist specifically for SNP data. For example, Genomic Control (GC) [9, 10] is an analytic method for SNP selection which controls the false positive rate by separating causal from confounding factors. There are also methods for selecting which SNPs to genotype when presented with a large number of arbitrary SNPs (see, for example, [37, 41]). However, we deemed such methods inappropriate for our context of interest in which we were presented with only the partial results of such methods, i.e., a modest number of SNPs not in linkage disequilibrium (LD) and without haplotype information which had been selected based upon the application of methods similar to those mentioned above (see Section 4 for more details on our context of interest).

## 3. Comparing and combining model predictions

Our goal is to determine whether the information from new binary covariates $X_j, j = 1, \ldots, p$, can be used to improve predictions of a response $Y$ from a model built on existing covariates $Z_l, l = 1, \ldots, p'$. Let $M_1$ and $M_2$ represent the logic regression and logistic regression models fit to $X_j, j = 1, \ldots, p$, respectively, and let $M_e$ represent the existing model fit to $Z_l, l = 1, \ldots, p'$. Suppose we are given a data set of size $n'$ consisting of covariates **X** and **Z** for which we would like to generate predicted values of the outcome $Y$. Let $\hat{Y}_1$ be the predictions for this data set from $M_1$, $\hat{Y}_2$ be the predictions from $M_2$, and $\hat{Y}_e$ be the predictions from $M_e$. Possible strategies for generating optimal predictions include the following:

- *Weighted Average of Predictions* $\bar{\hat{Y}}$. Determine whether a weighted average of the predictions from either $\hat{Y}_1$ or $\hat{Y}_2$ and $\hat{Y}_e$ yields better results than $\hat{Y}_e$ alone. A weighed average prediction $\bar{\hat{Y}}$ is defined as

$$(3.1) \qquad \bar{\hat{Y}} = \alpha \hat{Y}_e + (1-\alpha)\hat{Y}_m, \quad m = 1, 2,$$

  where $0 \leq \alpha \leq 1$. $\alpha$ is determined by repeated training/test set evaluation.
- *Predictions from Composite Model* $\hat{Y}_c$. We consider whether building a model directly to $\{\mathbf{X}, \mathbf{Z}\}$ will lead to improved predictions. The modeling procedures described in Section 2 are repeated with **Z** as well as **X** considered as possible covariates. This leads to models $M_{c1}$ (logic regression) and $M_{c2}$ (logistic regression) whose predictions $\hat{Y}_{c1}$ and $\hat{Y}_{c2}$ can be compared to $\hat{Y}_e$.
- *Two-Stage Predictions* $\hat{Y}_s$. Assume a two-class classification problem, i.e., $Y \in \{-1, 1\}$. In Stage 1 we determine for which observations the predictions $\hat{Y}_e$ are correct ($n_c \in \{1, \ldots, n'\}$) or incorrect ($n_{\bar{c}} = \{1, \ldots, n'\}/n_c$). In Stage 2 for observations in $n_{\bar{c}}$ we replace the predictions from $\hat{Y}_e$ with the predictions from either $M_1$ or $M_2$. In other words, we create a two-stage model $M_s$ for

$Y_i, i = 1, \ldots, n'$, with predictions defined as

$$\text{(3.2)} \qquad \hat{Y}_{si} = \begin{cases} \hat{Y}_{ei}, & \text{if } Y_{si} = Y_i, \\ \hat{Y}_{mi}, & \text{if } Y_{si} \neq Y_i, \end{cases}$$

where $m = 1, 2$. This predictive scheme may be particularly useful for data from a heterogenous population, where it is possible that the accuracy of the predictions from a given model may vary across different subgroups of the population.

Unfortunately $M_s$, and hence $\hat{Y}_s$, depends on the true response $Y$. As an alternative we propose the use of a support vector machine (SVM) to discriminate those subspaces of the covariate space on which the results of $M_e$ are correct from those on which the results are incorrect, based on the training data.

### 3.1. Support vector machines

Support vector machines (SVMs) [3, 8, 39] are a group of related supervised learning methods for classification or regression. In the case of two-class classification we consider a set of data points $\{(x_1, y_1), \ldots, (x_n, y_n)\}$ where each $y_i \in \{-1, 1\}$ denotes the class to which $x_i$ belongs. The objective of an SVM is to produce a hyperplane which can separate the two classes using only $x_i, i = 1, \ldots, n$ in a way which minimizes the empirical classification error and maximizes the geometric margin between the classes.

More specifically, the (soft margin) support vector machine is the solution to the following optimization problem:

$$\begin{array}{ll} \min_{w,b,\xi} & \frac{1}{2}w^t w + C \sum_{i=1}^{l} \xi_i, \quad C > 0, \\ \text{subject to} & y_i(w^t \phi(x_i) + b) \geq 1 - \xi_i, \quad \xi_i \geq 0. \end{array}$$

Note that the vectors $x_i, i = 1, \ldots, n$ are mapped to a higher dimensional space by the function $\phi$, and the SVM finds a linear separating hyperplane in this higher dimensional space. This SVM has a "soft margin" in the sense that is allows for misclassified samples; if no hyperplane exists which can separate the two classes, this method will chose the hyperplane which splits the classes as cleanly as possible while still maximizing the geometric margin. The slack variables $\xi_i$ measure the degree of misclassification of the datum $x_i$. $K(x_i, x_j) = \phi(x_i)^t \phi(x_j)$ is called the kernel function of the SVM. The kernel function typically falls into one of four classes: linear, polynomial, radial basis function, and sigmoid. For more information on SVMs and their implementation we refer the reader to [5, 31].

As stated previously, our use of SVMs is to discriminate those subspaces of the covariate space on which the results of the existing model $M_e$ are correct from those on which the results are incorrect, based on the training data. Let $\mathcal{X}$ be the covariate space and consider a SVM which divides $\mathcal{X}$ into subspaces $\mathcal{X}_c$ and $\mathcal{X}_{\bar{c}}$ where the model results are correct and incorrect, respectively. We now redefine the two-stage model $M_s$ (and $\hat{Y}_{si}$), originally defined in Equation (3.2), independently of $Y$ using the results of the SVMs:

$$\text{(3.3)} \qquad \hat{Y}_{si} = \begin{cases} \hat{Y}_{ei}, & \text{if } X_i \in \mathcal{X}_c, \\ \hat{Y}_{mi}, & \text{otherwise} \quad m = 1, 2. \end{cases}$$

*3.2. Analysis strategy recap*

Before we move to Section 4 we briefly summarize our analysis strategy. We want a statistical model which can accurately predict the outcome status of an observation given a set of existing predictors (both continuous and binary) and a set of new binary predictors. Our key modeling approach is a two-stage model. In the first stage we build a model from the existing predictors only (a logistic regression model). Given the predictions from this model we design an SVM which can identify the subspaces in which observations are correctly or incorrectly predicted. In the second stage, in those subspaces where observations are incorrectly predicted, we use a model based only on the new binary predictors (a logistic regression or logic tree model) to generate accurate predictions. In this approach information from the new binary predictors is only utilized where needed, i.e., in subspaces where the existing predictors do not provide enough information to generate accurate outcome predictions.

## 4. Example: The CATHGEN study

We demonstrate the use of our method in the analysis of data from a cardiology study. A substantial problem in clinical cardiology is the gap in the ability to detect asymptomatic individuals at high risk for coronary heart disease (CHD) for preventive and therapeutic interventions [17, 26]. Up to 75% of such individuals are designated as low to intermediate risk by standard CHD risk assessment models; however, a substantial number of such individuals who are actually at increased risk may not be identified. One analysis from the Framingham Heart Study found that for individuals that manifested a new CHD event, the initial presentation in over 50% of the cases was myocardial infarction, silent myocardial infarction or sudden cardiac death [1]. Over 50% of individuals with sudden cardiac death have no prior symptoms of CHD [40]. Therefore, it is likely that the traditional risk factors do not account fully for CHD risk [16, 22, 24, 28]. Furthermore, current CHD risk assessment models do not provide one's individual risk. Rather, the calculated assessment is for a population of individuals who share the same demographics and panel of risk factors.

A group of researchers at Duke University Medical Center (DUMC) has pursued an avenue of study evaluating the role of genes and gene variants in the development of atherosclerosis (the AGENDA study) [18, 19, 33]. As a result of their efforts they have compiled a list of candidate genes with a strong statistical correlation with vascular atherosclerosis. Through subsequent analysis for SNPs in these candidate genes, they analyzed 1300+ SNPs for association with significant CHD (stenosis $\geq$75% in at least one coronary artery) in a cohort of 1500 subjects who had undergone cardiac catheterization (CATHGEN). These SNPs were then ranked by their marginal association with the presence of CHD in a cardiac catheterization population.

We conduct an analysis of a subset of the CATHGEN data to test the hypothesis that genetic information in the form of SNPs will improve the ability of risk assessment models that use only traditional risk factors to classify individuals as having high risk for CHD. We developed prediction models for likelihood of significant CHD based on traditional risk factors such as cholesterol, blood pressure, diabetes and smoking, using a group of CATHGEN subjects who underwent cardiac catheterization. A separate set of CATHGEN subject data was used in selecting from the candidate SNP pool those SNPs with the highest marginal association with significant CHD; 81 such SNPs were available for analysis. We then assessed whether including genetic information improved our ability to classify individuals as having significant CHD.

The research was performed under an approved protocol from the Institutional Review Board of DUMC.

## 4.1. Data

Two data sets were constructed from the CATHGEN data, one for SNP selection and model building (build set) and one for the evaluation of model predictions (evaluation or eval set). The evaluation set consisted of white individuals (self-reported race) with complete data for all 81 SNPs and all clinical variables (see Section 4.2). The build set consisted of white individuals with complete data for all 81 SNPs but incomplete clinical data (clinical data was assumed to be missing at random). Within each set individuals were separated into three cohorts: a) controls, $\geq 65$ years of age without significant CHD, b) older cases (OC), $\geq 65$ years of age with significant CHD, and c) younger cases (YC), $\leq 50$ years of age with significant CHD. For each cohort a group of samples was selected for model validation only, those 50–55 years of age with either minimal or significant CHD as defined by coronary angiography (for validation of models of cohorts a) and c)) and those 56–65 years of age with either minimal or significant CHD as defined by coronary angiography (for validation of models of cohorts a) and b)). Each cohort was further split by gender. A table of the study cohorts and number of subjects is shown in Table 1.

We used the 81 SNPs from the AGENDA study with the strongest statistical association with the presence of significant CHD. The strength of association was determined by 1) the p-value of SNP status in a logistic regression model of CHD, including both age and gender as covariates, and 2) the p-value of SNP status from a Cochran-Armitage Test for Trend [2]. We should note that the designation of the top SNPs was performed using a large group of subjects (1500) that included the data used for this study.

Typically with any given single base-pair difference, or single nucleotide polymorphism (SNP), only two out of the four possible nucleotides occur. Since each cell contains a pair of every autosome, we can think of a SNP as a three-level variable $X$ taking the values 0, 1, or 2 (e.g., for nucleotide pairs A/A, A/G, and G/G, respectively). Each SNP can be recoded as a binary variable using either dominant coding ($X_d = 1$ if $X \geq 1$ and $X_d = 0$ otherwise) or recessive coding ($X_r = 1$ if $X = 2$ and $X_r = 0$ otherwise). With the CATHGEN data we chose dominant coding for the SNPs, where $X = 0$ indicates no copy of the minor (less frequently occurring) allele.

TABLE 1
*CATHGEN data*

|  |  | Build set | | Evaluation set | |
|---|---|---|---|---|---|
|  |  | **Training** | **Validation** | **Training** | **Validation** |
| Male | Young Cases | 44 | 80 | 69 | 103 |
|  | Controls | 34 | 14 | 32 | 18 |
|  | Older Cases | 79 | 13 | 47 | 11 |
| Female | Young Cases | 11 | 21 | 11 | 18 |
|  | Controls | 59 | 12 | 42 | 18 |
|  | Older Cases | 15 | 3 | 15 | 4 |

## 4.2. Model building

Models were constructed on either male or female subjects. Within gender, these models compared either controls with young cases or controls with older cases. We describe the modeling approaches used for each gender/comparison combination. All computations were performed in R [27].

*Predictive model using clinical variables* ($M_e$).

For the clinical variables we used standard CHD risk factors as denoted by assessment tools such as the Framingham heart study risk algorithm [40]. We included presence of diabetes, current smoking status, total cholesterol level, HDL cholesterol level, systolic blood pressure and diastolic blood pressure. Clinical variables were collected at the time of cardiac catheterization. We used these variables to train both weighted and unweighted logistic regression models in the evaluation set, as the build set has incomplete clinical data. The weights were chosen to balance the importance of case and control samples. The trained model was then used to classify the validation subjects in the evaluation set as having minimal or significant CHD.

*Predictive model using genetic variables* ($M_1, M_2$).

Our SNP data consisted of the 81 SNPs from the AGENDA study typed in our CATHGEN samples, as described in Section 4.1. We constructed two models using only the CATHGEN samples from the build set: 1) LASSO for SNP selection followed by logistic regression (weighted and unweighted) using backwards selection, and 2) logic regression based on all 81 SNPs. Logic models were fit for both classification and logistic regression. These models were then used to classify the subjects in the evaluation set as having minimal or significant CHD. LASSO and logic regression were performed using the R packages `lasso2` and `LogicReg`, respectively.

*Predictive model using combined clinical and genetic variables* ($M_c$).

First, logistic regression models (weighted and unweighted) were built using genetic variables, as described above. The SNPs which appear in each model and the clinical variables were combined to train logistic regression models in the evaluation set. These models were then used to classify the validation subjects in the evaluation set as having minimal or significant CHD. A similar procedure was performed for each logic regression model.

*Two-Stage Predictive model using the clinical and genetic models* ($M_s$).

First, the trained clinical model was used to classify the subjects in the evaluation set as having minimal or significant CHD. Next an SVM was constructed which could discriminate the subspace of correctly classified samples from the subspace of incorrectly classified samples. For those samples in the subspace of incorrectly classified samples, the trained genetic models were applied to reclassify the subjects into the minimal and significant CHD groups. This resulted in a set of two-stage predictions, as described in Section 3.1 and Equation (3.3). SVMs were based on a radial basis function kernel and computed using the R package `e1071` [12, 23].

## 4.3. Results

Our interest is in classifying individuals as having non-significant CHD ($Y = 0$) or significant CHD ($Y = 1$). A fitted or predicted probability of significant CHD $\hat{Y}_i$

Ensembles for prediction311Ensembles for prediction      311
TABLE 2
*Results of clinical model and logic regression classification models on evaluation data*

|  | Training/Test Samples | | | | Validation Samples | | | |
| --- | --- | --- | --- | --- | --- | --- | --- | --- |
|  | $\hat{Y}_e$ | $\hat{Y}_1$ | $\hat{Y}_c$ | $\hat{Y}_s$ | $\hat{Y}_e$ | $\hat{Y}_1$ | $\hat{Y}_c$ | $\hat{Y}_s$ |
| Female, controls vs older cases | | | | | | | | |
| auROC | 71.38 | 48.81 | 75.38 | **81.31** | 50.48 | 57.62 | 53.33 | **75.24** |
| acc | 71.93 | 50.88 | 66.67 | **82.46** | 50.00 | 68.18 | 54.55 | **81.82** |
| fn | 36.00 | 68.00 | 40.00 | **28.00** | 57.14 | 71.43 | **42.86** | **42.86** |
| fp | 21.88 | 34.38 | 28.13 | **9.38** | 46.67 | 13.33 | 46.67 | **6.67** |
| Male, controls vs older cases | | | | | | | | |
| auROC | 81.97 | 51.46 | **82.97** | 74.18 | 59.52 | 42.86 | 59.94 | **60.00** |
| acc | 78.48 | 73.42 | 78.48 | **87.34** | 51.72 | 41.38 | 83.47 | **58.62** |
| fn | 11.48 | 8.20 | 11.48 | **1.64** | 14.29 | 14.29 | 5.66 | **0.00** |
| fp | 55.56 | 88.89 | 55.56 | **50.00** | **80.00** | 100.0 | 93.33 | **80.00** |
| Female, controls vs younger cases | | | | | | | | |
| auROC | 80.80 | 50.15 | 80.80 | **86.53** | 63.81 | 50.00 | 63.81 | **85.71** |
| acc | 79.25 | 56.60 | 79.25 | **88.68** | 61.11 | 47.22 | 61.11 | **83.33** |
| fn | 33.33 | 80.95 | 33.33 | **23.81** | 42.86 | 66.67 | 42.59 | **28.57** |
| fp | 12.50 | 18.75 | 12.50 | **3.13** | 33.33 | 33.33 | 33.33 | **0.00** |
| Male, controls vs younger cases | | | | | | | | |
| auROC | 84.00 | 37.01 | **86.81** | 73.19 | 60.19 | 58.33 | 59.94 | **88.11** |
| acc | 83.17 | 46.53 | 83.17 | **88.12** | 81.82 | 52.07 | 83.47 | **94.21** |
| fn | 4.82 | 48.19 | 4.82 | **3.61** | 8.49 | 50.00 | 5.66 | **3.77** |
| fp | 72.22 | 77.78 | 72.22 | **50.00** | 86.67 | 33.33 | 93.33 | **20.00** |

from a logistic regression model for a given individual was considered an "accurate" classification if $Y_i = 1$ and $\hat{Y}_i \geq 0.5$, or $Y_i = 0$ and $\hat{Y}_i < 0.5$, and considered "inaccurate" otherwise. A fitted or predicted outcome $\hat{Y}_i$ from a logic regression model for a given individual was considered "accurate" if $Y_i = \hat{Y}_i$ and considered "inaccurate" otherwise. We considered overall model accuracy as well as the rate of false positive ($P(\hat{Y}_i \geq 0.5 | Y_i = 0)$) and false negative ($P(\hat{Y}_i < 0.5 | Y_i = 1)$) model results. In an attempt to balance specificity and sensitivity we also calculated the area under the receiver-operating characteristic (ROC) curve (auROC) for the results of each model. The auROC is equal to the value of the Wilcoxon–Mann–Whitney statistic and can be interpreted as the probability that the model will assign a higher probability of significant CHD to a randomly selected positive sample than to a randomly selected negative sample. The auROC calculations were performed with the R package ROCR [34].

The results of the weighted logistic regression models and the logic regression classification and logistic models for each gender and comparison (control vs. young cases or control vs. older cases) on the evaluation set are presented in Tables 2, 3, and 4. The results of the unweighted logistic regression models are not discussed here due to their similarity to the results of the weighted models. In these Tables $\hat{Y}_e$ = clinical only model, $\hat{Y}_1$ or $\hat{Y}_2$ = SNP only model, $\hat{Y}_c$ = Clinical+SNP model, $\hat{Y}_s$ = Two-Stage Predictions using SVM, acc = accuracy, fn = false negative rate, and fp = false positive rate.

These tables show clearly that the two-stage predictions yield the best results. In some cases the combined clinical+SNP models perform better on the training set in comparison to the two-stage predictions, but their performance deteriorates on the validation samples. This could be due to the fact that the clinical model and the clinical+SNP model were trained on the training/test samples and tested on the validation samples, while the SNP models were tested on both sets of samples (having already been trained on the samples in the build set). This would also explain the consistency of the SNP model results across both the training/test and validation samples.

TABLE 3
*Results of clinical model and logic regression logistic models on evaluation data*

|  | Training/Test Samples |  |  |  | Validation Samples |  |  |  |
|---|---|---|---|---|---|---|---|---|
|  | $\hat{Y_e}$ | $\hat{Y_1}$ | $\hat{Y_c}$ | $\hat{Y_s}$ | $\hat{Y_e}$ | $\hat{Y_1}$ | $\hat{Y_c}$ | $\hat{Y_s}$ |
| Female, controls vs older cases |  |  |  |  |  |  |  |  |
| auROC | 71.38 | 52.50 | 72.50 | **84.63** | 50.48 | 59.05 | 48.57 | **79.53** |
| acc | 71.93 | 49.12 | 63.16 | **84.21** | 50.00 | 54.55 | 31.82 | **72.73** |
| fn | 36.00 | 20.00 | 44.00 | **12.00** | 57.14 | 28.57 | 42.86 | **14.29** |
| fp | 21.88 | 75.00 | 31.25 | **18.75** | 46.67 | 53.33 | 80.00 | **26.67** |
| Male, controls vs older cases |  |  |  |  |  |  |  |  |
| auROC | 81.97 | 59.38 | 82.97 | **88.39** | 59.52 | 43.81 | 63.33 | **82.86** |
| acc | 78.48 | 49.36 | 82.28 | **91.14** | 51.72 | 44.83 | 51.72 | **82.76** |
| fn | 11.48 | 59.02 | **4.92** | 6.56 | **14.29** | 85.71 | **14.29** | 14.29 |
| fp | 55.56 | 22.22 | 61.11 | **16.67** | 80.00 | 26.67 | 80.00 | **20.00** |
| Female, controls vs younger cases |  |  |  |  |  |  |  |  |
| auROC | 80.80 | 52.60 | **81.40** | 80.21 | 63.81 | 50.00 | 64.13 | **80.95** |
| acc | 79.25 | 56.60 | 70.70 | **83.02** | 61.11 | 47.22 | 55.56 | **77.78** |
| fn | **33.33** | 66.67 | 47.62 | **33.33** | 42.86 | 66.67 | 47.62 | **38.10** |
| fp | 12.50 | 28.13 | 15.63 | **6.25** | 33.33 | 33.33 | 40.00 | **0.00** |
| Male, controls vs younger cases |  |  |  |  |  |  |  |  |
| auROC | 84.00 | 47.52 | **91.43** | 77.78 | 60.19 | 57.83 | 61.32 | **80.97** |
| acc | 83.17 | 49.50 | 81.13 | **92.08** | 81.82 | 57.20 | 78.51 | **91.74** |
| fn | 4.82 | 49.40 | 4.82 | **0.00** | 8.49 | 44.34 | 12.26 | **4.72** |
| fp | 72.22 | 55.56 | 50.00 | **44.44** | 86.67 | 40.00 | 86.67 | **33.33** |

TABLE 4
*Results of clinical model and weighted logistic regression models on evaluation data*

|  | Training/Test Samples |  |  |  | Validation Samples |  |  |  |
|---|---|---|---|---|---|---|---|---|
|  | $\hat{Y_e}$ | $\hat{Y_1}$ | $\hat{Y_c}$ | $\hat{Y_s}$ | $\hat{Y_e}$ | $\hat{Y_1}$ | $\hat{Y_c}$ | $\hat{Y_s}$ |
| Female, controls vs older cases |  |  |  |  |  |  |  |  |
| auROC | 71.38 | 54.56 | 77.38 | **84.00** | 50.48 | 60.48 | 49.52 | **75.24** |
| acc | 71.93 | 56.14 | 70.18 | **85.96** | 50.00 | 72.73 | 50.00 | **81.82** |
| fn | 36.00 | 64.00 | **32.00** | **32.00** | 57.14 | 71.43 | **42.86** | **42.86** |
| fp | 21.88 | 28.13 | 28.13 | **0.00** | 46.67 | **6.67** | 53.33 | **6.67** |
| Male, controls vs older cases |  |  |  |  |  |  |  |  |
| auROC | 81.97 | 47.27 | **93.62** | 77.78 | 59.52 | 45.00 | 66.19 | **73.33** |
| acc | 78.48 | 65.82 | 88.61 | **89.87** | 51.72 | 51.72 | 48.28 | **72.41** |
| fn | 11.48 | 22.95 | 6.56 | **0.00** | 14.29 | 35.71 | 21.43 | **0.00** |
| fp | 55.56 | 72.22 | **27.78** | 44.44 | 80.00 | 60.00 | 80.00 | **53.33** |
| Female, controls vs younger cases |  |  |  |  |  |  |  |  |
| auROC | 80.80 | 45.47 | **83.33** | 81.77 | 63.81 | 52.38 | 49.52 | **78.57** |
| acc | 79.25 | 56.60 | 75.47 | **84.91** | 61.11 | 38.89 | 50.00 | **75.00** |
| fn | **33.33** | 100.0 | **33.33** | **33.33** | **42.86** | 95.24 | **42.86** | **42.86** |
| fp | 12.50 | 6.25 | 18.75 | **3.13** | 33.33 | 13.33 | 53.33 | **0.00** |
| Male, controls vs younger cases |  |  |  |  |  |  |  |  |
| auROC | 84.00 | 51.94 | **95.79** | 82.13 | 60.19 | 62.52 | 66.19 | **91.45** |
| acc | 83.17 | 47.62 | **93.07** | 92.08 | 81.82 | 50.41 | 48.28 | **95.04** |
| fn | 4.82 | 54.22 | **2.41** | **2.41** | 8.49 | 52.83 | 21.43 | **3.77** |
| fp | 72.22 | 44.44 | **27.78** | 33.33 | 86.67 | 26.67 | 80.00 | **13.33** |

Interestingly the combined clinical+SNP models did not perform better than the clinical only or SNP only models on the validation samples. In many comparisons the clinical and clinical+SNP models performed comparably, with the SNP models performing quite poorly. We surmise that the population under study is quite heterogeneous, and that no one data type provides information predictive of CHD in all subpopulations. The clinical data is predictive for some samples while the genetic data is predictive for others. The results of a weighted average of the predictions from the clinical only and SNP only models ($\bar{\hat{Y}}$; results not shown) were not successful because both data types are not relevant for all samples; often the data types provide conflicting information. The two-stage predictions are an attempt to use an SVM to define subpopulations for which the clinical data or the genetic data are predictive. Using the SVM results we can identify which data type is predictive for a given sample, leading to more accurate predictions overall.

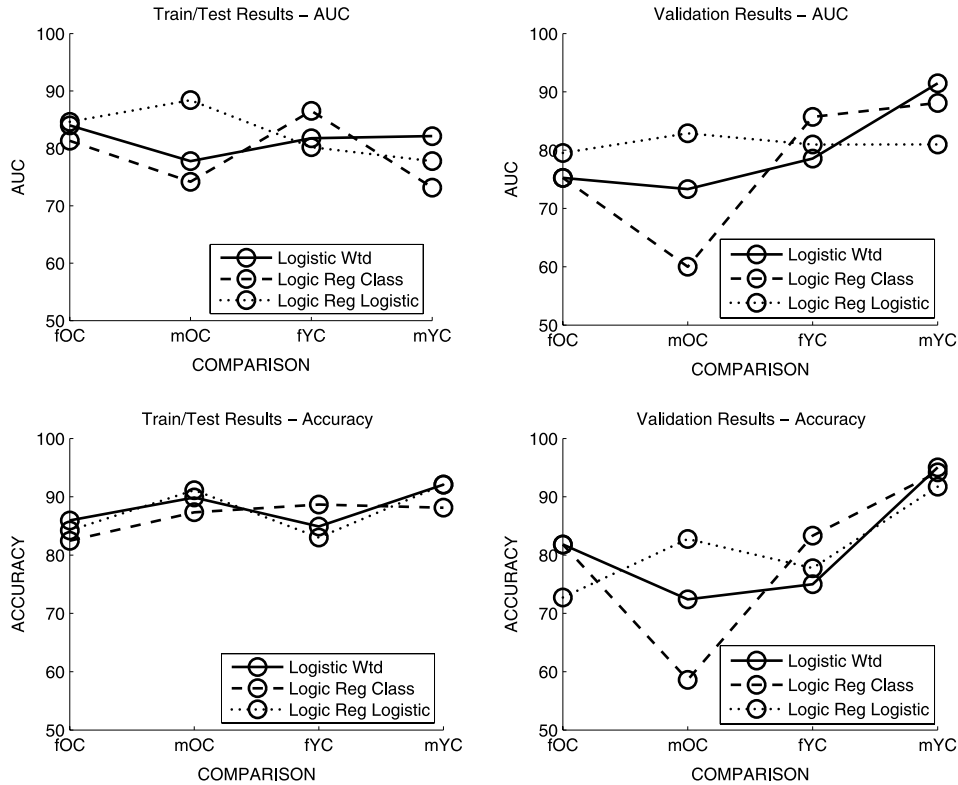

Fig 2. *Quality of two-stage predictions on evaluation data.*

In Figure 2 we display the quality of the two-stage predictions on the evaluation set (train/test subset or validation subset) for each model and each comparison. No single model performs consistently best in all comparisons. By averaging the performance measures (auROC or accuracy) on the validation samples across comparisons we find in terms of auROC the logic regression logistic models perform 1.43% better than the weighted logistic regression models, which in turn perform 2.38% better than the logic regression classification models. In terms of accuracy the logic regression logistic models perform 0.19% better than the weighted logistic regression models, which in turn perform 1.57% better than the logic regression classification models. Hence we conclude that overall the logic regression logistic models perform best, followed by the weighted logistic regression models and finally the logic regression classification models. However, the difference in average performance between any two model types is quite small.

The only comparison in which model performances are clearly distinguished is the male controls vs. older cases comparison. The relatively poor performance of the two-stage predictions from the logic regression classification models is striking; the results in Table 2 reveal that the logic regression classification model had false positive rates of 89% for the training/test patients and 100% for the validation patients. Unfortunately the clinical model also had a high false positive rate of 80%. Both data types (and consequently the two-stage predictions) failed to predict those with minimal CHD. This is possibly a result of SNP selection; of the 9, 8, and 8 SNPs selected by the weighted logistic, logic regression classification and logistic models, respectively, only 2 appear in all three models and no other SNPs are shared by any two models. No definitive conclusions can be drawn without an independent data set on which to validate our results.

## 5. Discussion

We have presented a two-stage approach to generating combined predictions from models built from different data sources. One model is built on existing data of multiple types (e.g., traditional clinical risk factors), while a second set of models are built on newly available binary predictors only (in our case genetic SNP data). This two-stage approach uses an SVM to distinguish the covariate subspaces on which the existing data model generates accurate or inaccurate predictions. The existing model is used to generate predictions for samples in the "accurate" subspace while a model built on the newly available data is used to generate predictions for samples in the "inaccurate" subspace. This approach appears to perform well in generating predictions for a heterogeneous population for which no single data type provides predictive information for all samples.

As discussed briefly in Section 1 there exist modeling approaches other than logistic and logic regression models which could have been employed here. We chose logic trees because of their ability to capture higher order interactions, an issue of great importance in regression and a key to variable selection. However, similar models could be constructed by Bayesian model averaging with lower-dimensional logistic regression models that allow for interactions among covariates. We also could have employed neural networks or projection pursuit models. These alternative approaches would require careful prior variable selection in any context where $n < p$, but would be worth considering in future work.

Our modeling approach is similar in spirit to ensemble methods [11], learning algorithms which construct a set of classifiers and then generate predictions by taking a (weighted) average or vote of their predictions. One such approach is boosting [13, 14, 32], a method for converting a weak learning algorithm into one with high accuracy. This is done by training classifiers on weighted versions of the training data, giving higher weight to misclassified samples, and forming the final classifier as a linear combination of the training classifiers. This approach does not apply different models to different covariate subspaces, but does attempt to improve model performance in subspaces where the model performs poorly. Our approach is a type of ensemble method in which each classifier gets either a single, fully weighted vote or no vote depending upon the subspace in which the sample of interest is located. It would be of interest to compare our two-stage predictive approach to an approach aimed at building a boosted classifier from all available covariates. The results of such a comparison would help in determining the necessity of building a subspace-dependent classifier.

Several different model types were used in generating predictions from the newly available binary data, including logistic regression and logic regression models. No single model type performed significantly better than the others, although a slight performance advantage was observed when using the two-stage predictions from the logic regression logistic models. Across comparisons within gender and case the best models generated two-stage predictions with validation accuracies between 81.82% and 94.21%. It should be noted, however, that the sizes of the validation sets for some comparisons are quite small and all comparisons were conducted within a single population (CATHGEN). Also, our inferences are done conditional on a fixed chosen model; the variability of the models is not considered in the inference procedure. This is a weakness in our approach as model uncertainty can be substantial in high dimensional data contexts. Hence we regard our results as a "proof-of-concept" for our analysis approach. We are planning an analysis of a second, independent population and await the results of such an analysis before making any definitive conclusions regarding the predictive power of our method.

**Acknowledgments.** The authors wish to thank the following for their assistance: Bertrand Clarke, Department of Statistics, University of British Columbia; Ed Iversen, Department of Statistical Science, Duke University; Pascal Goldschmidt, Dean, Leonard M. Miller School of Medicine, University of Miami.

316                                J. Clarke and D. Seo[14] Freund, Y. and Schapire, R. (1997). A decision-theoretic generalization of on-line learning and an application to boosting. *J. Comput. System Sci.* **55** 119–139. MR1473055
[15] Friedman, J. (1991). Multivariate adaptive regression splines (with discussion). *Ann. Statist.* **19** 1–141. MR1091842
[16] Greenland, P., Knoll, M., Stamler, J., Neaton, J., Dyer, A., Garside, D. and Wilson, P. (2003). Major risk factors as antecedents of fatal and nonfatal coronary heart disease events. *J. Amer. Medical Association* **290** 891–897.
[17] Greenland, P., Smith, S. and Grundy, S. (2001). Improving coronary heart disease risk assessment in asymptomatic people: Role of traditional risk factors and noninvasive cardiovascular tests. *Circulation* **104** 1863–1867.
[18] Hauser, E., Crossman, D., Granger, C., Haines, J., Jones, C., Mooser, V., McAdam, B., Winkelmann, B., Wiseman, A., Muhlstein, J., Bartel, A., Dennis, C., Dowdy, E., Estabrooks, S., Eggleston, K., Francis, S., Roche, K., Clevenger, P., Huang, L., Pedersen, B., Shah, S., Schmidt, S., Haynes, C., West, S., Asper, D., Booze, M., Sharma, S., Sundseth, S., Middleton, L., Roses, A., Hauser, M., Vance, J., Pericak-Vance, M. and Kraus, W. (2004). A genomewide scan for early-onset coronary artery disease in 438 families: The GENECARD study. *Amer. J. Human Genetics* **75** 436–447.
[19] Karra, R., Vermullapalli, S., Dong, C., Herderick, E., Song, X., Slosek, K., Nevins, J., West, M., Goldschmidt-Clermont, P. and Seo, D. (2005). Molecular evidence for arterial repair in atherosclerosis. *Proc. Nat. Acad. Sci. U.S.A.* **102** 16789–16794.
[20] Kooperberg, C., Ruczinski, I., LeBlanc, M. and Hsu, L. (2001). Sequence analysis using logic regression. *Genetic Epidemiology* **21** S626–S631.
[21] Lokhorst, J., Venables, B., Turlach, B. and Maechler, M. (2006). The lasso2 package: L1 constrained estimation aka "lasso." Univ. Western Australia School of Mathematics and Statistics. Version 1.2-5. Available at http://www.maths.uwa.edu.au/~berwin/software/lasso.html.
[22] Magnus, P. and Beaglehole, R. (2001). The real contribution of the major risk factors to the coronary epidemics: time to end the "only-50%" myth. *Archives of Internal Medicine* **161** 2657–2660.
[23] Meyer, D. (2006). Support vector machines: The interface to libsvm in package e1071. Technische Universität Wien, Austria.
[24] Mosca, L. (2002). C-Reactive protein: To screen or not to screen? *New England J. Medicine* **347** 1615–1617.
[25] Osborne, M., Presnell, B. and Turlach, B. (2000). On the LASSO and its dual. *J. Comput. Graph. Statist.* **9** 319–337. MR1822089
[26] Pasternak, R., Abrams, J., Greenland, P., Smaha, L., Wilson, P. and Houston-Miller, N. (2003). Task force #1 – identification of coronary heart disease risk – is there a detection gap? *J. American College of Cardiology* **41** 1863–1874.
[27] R Development Core Team (2006). R: A language and environment for statistical computing. R Foundation for Statistical Computing, Vienna, Austria. Available at http://www.R-project.org.
[28] Ridker, P., Rifai, N., Rose, L., Buring, J. and Cook, N. (2002). Comparison of C-reactive protein and low-density lipoprotein cholesterol levels in the prediction of first cardiovascular events. *New England J. Medicine* **347** 1557–1565.